\documentclass[12pt]{article}
\usepackage{epsfig}

\newcommand\pubnumber{UCTP-111(00)}
\newcommand\pubdate{\today}

\def\napoli{Department of Physics\\
University of Cincinnati, Cincinnati, OH 45221-0011, USA \\
(For The Belle Collaboration)}
\def\support{\footnote{email : satpathy@bsunsrv1.kek.jp}}

\def\Title#1{\begin{center} {\Large #1 } \end{center}}
\def\Author#1{\begin{center}{ \sc #1} \end{center}}
\def\Address#1{\begin{center}{ \it #1} \end{center}}

\newcommand\pubblock{\rightline{\begin{tabular}{l} \pubnumber\\
         \pubdate  \end{tabular}}}
\newenvironment{Abstract}{\begin{quotation}  }{\end{quotation}}
\newenvironment{Presented}{\begin{quotation} \begin{center} 
             Presented at the\end{center}\bigskip 
      \begin{center}\begin{large}}{\end{large}\end{center} \end{quotation}}
\def\Acknowledgements{\bigskip  \bigskip \begin{center} \begin{large}
             \bf ACKNOWLEDGMENTS \end{large}\end{center}}

\def\beq{\begin{equation}}
\def\eeq#1{\label{#1}\end{equation}}
\def\eeqn{\end{equation}}

\def\beqa{\begin{eqnarray}}
\def\eeqa#1{\label{#1}\end{eqnarray}}
\def\eeqan{\end{eqnarray}}

\let\bar=\overbar

\def\Dslash{\not{\hbox{\kern-4pt $D$}}}
\def\dslash{\not{\hbox{\kern-2pt $\del$}}}

\def\msb{{\bar{\ssstyle M \kern -1pt S}}}

\begin{document}
\begin{titlepage}
\pubblock

\vfill
\Title{First Physics Results From Belle}
\vfill
\Author{Asish Satpathy\support}
\Address{\napoli}
\vfill
\begin{Abstract}
The Belle detector at the KEK-B asymmetric $e^{+}e^{-}$ collider has 
recorded 6.2 fb$^{-1}$ data at the $\Upsilon(4S)$ resonance by July 2000. 
Using this data sample, several new results on 
various B meson branching ratio measurements are presented.
We also report on the measurement of the Standard Model $CP$ 
violation parameter $\sin(2\phi_{1})$, where $\phi_{1}$ 
is one of angles of the CKM triangle. The preliminary result is 	
$\sin(2\phi_{1})$ = $0.45^{+0.43 \pm 0.07}_{-0.44 \pm 0.08}$.

\end{Abstract}
\vfill
\begin{Presented}
5th International Symposium on Radiative Corrections \\ 
(RADCOR--2000) \\[4pt]
Carmel CA, USA, 11--15 September, 2000
\end{Presented}
\vfill
\end{titlepage}
\def\thefootnote{\fnsymbol{footnote}}
\setcounter{footnote}{0}

\section{Introduction}
The Belle Experiment ~\cite{belle_detector}
at the KEK-B asymmetric $e^{+}~e^{-}$ collider has completed its
first year of operation in July 2000 accumulating data equivalent to an integrated luminosity of 6.2 fb$^{-1}$ on the $\Upsilon(4S)$. This corresponds 
to about 6.3 $\times 10^6 B\bar{B}$'s. 
Apart from the early running period, 
the KEK-B machine operated quite well delivering a record luminosity of 
94 pb$^{-1}$ per day and 504 pb$^{-1}$ per week respectively
and has great prospects for further improvement towards its design goal of 
100 fb$^{-1}$ / Year. We report on some of the new measurements that
have been carried out using this dataset with the emphasis on the measurement of 
the Standard Model CP violation parameter $\sin(2\phi_{1})$. The results
being reported are all preliminary.

\section{General Features of Data Analysis}

$B$ candidates are identified using 
$M_{b} = \sqrt{E^{2}_{beam} - |\sum P_{i}^{cms}|^{2}}$, the beam 
constrained mass and 
$\Delta E = E_{beam}-E_{B}$, where $E_{beam} = E_{cms}/2$.
While $M_{b}$ expresses the momentum conservation in the decay,
$\Delta E$ expresses the energy conservation of the particles in the decay
and is sensitive to the missing particles and $K/\pi$ misidentification. 

In all the decay modes we consider here, 
the dominant source of background arises from $e^{+}e^{-} \rightarrow q\bar{q}
(q=u,d,c,s)$ transitions. We exploit the difference between jetty 
hadronization of continuum events and spherical decay of $B$'s at 
$\Upsilon(4S)$ center of mass frame. Continuum background is reduced 
using $R_{2}$ = $H_{2}$/$H_{0}$, where  
$H_{l} = \sum_{i,j} \frac{|\vec{p_{i}}||\vec{p_{j}}|}{E_{cms}^{2}} P_{l}
(\cos \theta_{i,j})$, the Fox-Wolfram moment ~\cite{Fox}, that measures 
the shape of an event as a whole and $P_{l}(\cos \theta_{i,j})$ 
are Legendre Polynomials.
In some decay modes, the continuum background is
reduced using modified Fox Wolfram moments ~\cite{SFW} where the tracks 
and showers coming from $B$ and the rest of the tracks and showers
in the event are separated. 
These modified Fox-Wolfram moments are combined in a Fisher Discriminant to
form the Super Fox Wolfram (SFW).
The SFW has approximately 2$\sigma$ separation between $q\bar{q}$ and
$B\bar{B}$ events and provides a 22 \% increase in the expected significance
for some modes as compared to $R_{2}$.
Another popular technique to remove the continuum 
component in the data is to cut on $\cos(\theta_{thrust})$ variable,
where $\theta_{thrust}$ is the angle between 
the thrust axis of the signal $B$ and the thrust axis of the rest of the event.
In some analyses, the continuum variables are combined with kinematic variables
in a likelihood fit to determine the signal yield.

\section{Branching Ratio Measurement}
We will highlight selected branching ratio measurements that
were either first observation or important new 
measurements. The preliminary results reported in this section are
based on the first 5.3 million $B\bar{B}$ events recorded 
on the $\Upsilon(4S)$ resonance at KEK-B.

\subsection{$B \rightarrow \phi K$ }
\begin{figure}[htb]
\begin{center}
\epsfig{file=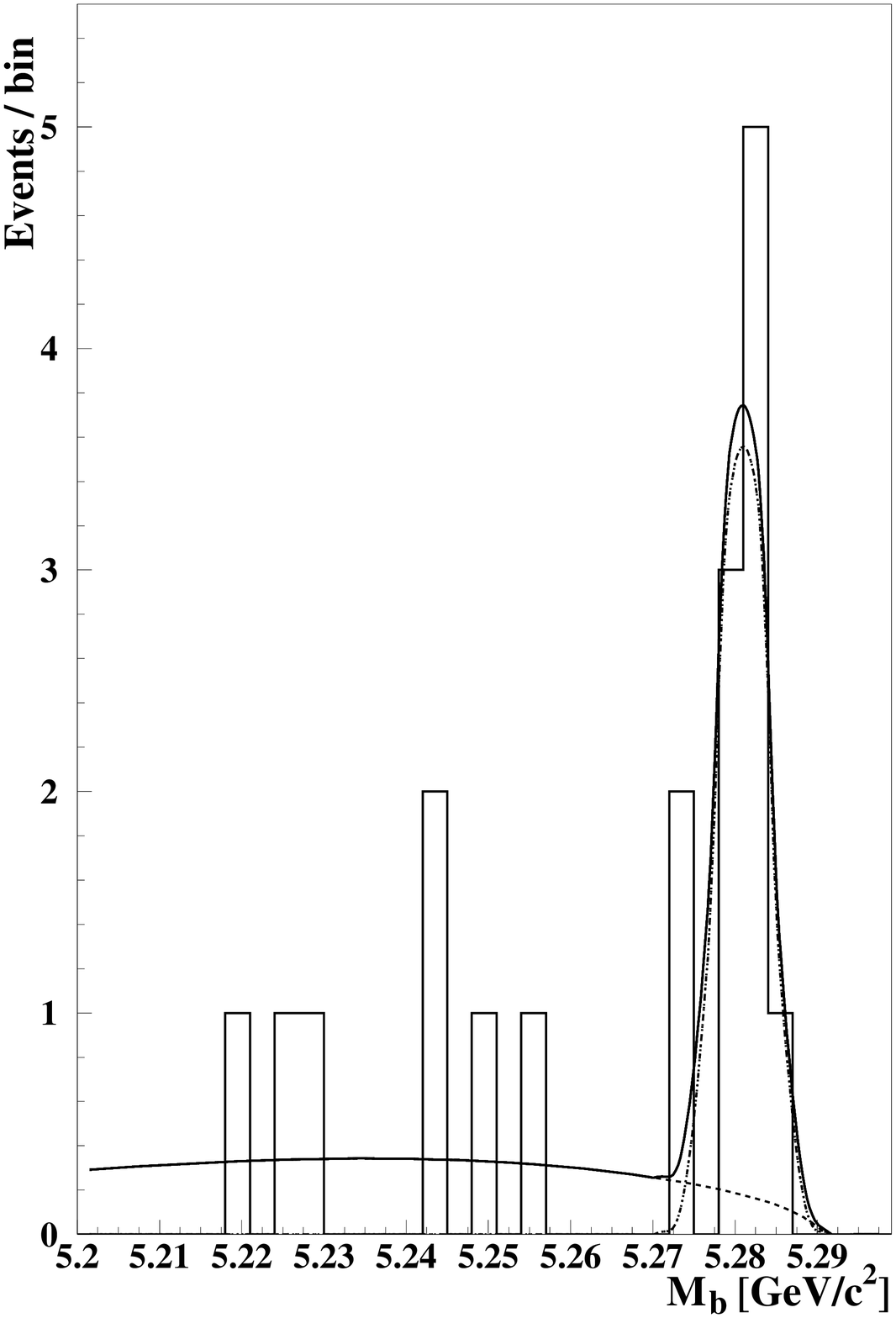,height=2.0in, width=2.5in}
\epsfig{file=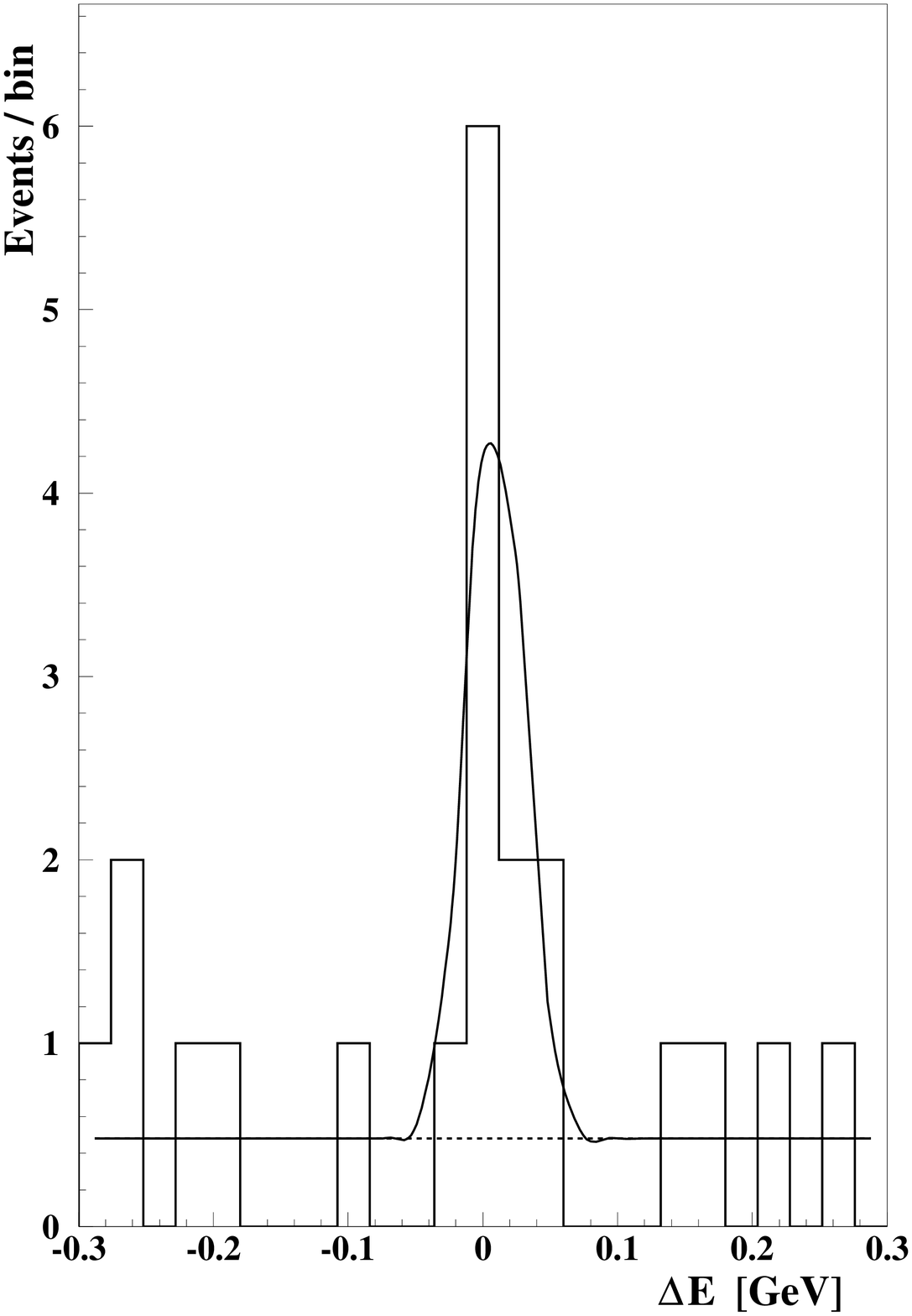,height=2.0in, width=2.5in}
\caption{$B \rightarrow \phi K$ signal yield. [left:] Beam constrained
mass distribution [right:] $\Delta E$ distribution }
\label{fig:phiK}
\end{center}
\end{figure}

This is the first observation of $B$ decays involving a pure penguin transition,
$b \rightarrow s s \bar{s}$.
Two charged tracks identified as kaons are combined to form a $\phi$ meson candidate
with an additional requirement that both tracks are from one vertex.
The candidate $\phi$ is then combined with a charged $K$ or $K_{s}$ to form
a $B$ candidate. Continuum background is suppressed with cuts on 
$\cos(\theta_{thrust})$, the $\phi$ meson helicity angle which 
is the angle between the direction of $K^+$ and the momentum vector
of $\phi(1020)$ and the $B$ flight 
direction. The final yield is obtained by fitting the $M_{b}$ distribution 
(Fig.~\ref{fig:phiK}). The binned likelihood fit yields $9.2_{-2.9}^{+3.6}$ 
($B \rightarrow \phi K$) events with a statistical significance of 5.4$\sigma$ 
and ${\cal B} (B^{+} \rightarrow \phi(1020) K^{\pm})
= 1.72_{-0.54}^{+0.67}$(stat.) $\pm 0.18$(sys.) $\times 10^{-5}$.
We also observed two events of the type $B \rightarrow \phi K_{s}$ in the 
3$\sigma$
signal box. These are consistent with the fluctuation in the $q\bar{q}$ background.
A more detailed description of the analysis can be found in  
reference ~\cite{b2phik}.
As statistics improve, this decay mode will be used
to determine the CKM angle $\phi_{1}$.

\subsection{$B \rightarrow K\pi, KK, \pi\pi$}
The study of charmless hadronic $B$ meson decays offers a variety of 
test of Standard Model physics and beyond. Our immediate motivation was to 
measure the branching fraction of those decay modes which are
either not measured or are limited by statistics.
We have summarized the current Belle branching fraction 
measurements of charmless $B \rightarrow h h$ decays in Table~\ref{tab:B2PP}.
Thanks to the excellent performance of the high momentum particle identification
system, we have seen very clear signals in some of the decay modes.
A detailed account of the analysis can be found in references ~\cite{b2pp1}
and ~\cite{b2pp2}.
The measurements are all consistent with similar observations in other
experiments~\cite{other}.

\begin{table}[t]
\begin{center}
\begin{tabular}{llll}  
Decay Modes & Signal Yield  & ${\cal B}(\times 10^{-5})$ 
& U.L($\times 10^{-5}$) \\ \hline

$B^{0} \rightarrow K^{+} \pi^{-}$ & $25.6_{-6.8}^{+7.5}$ 
& $1.74_{-0.46}^{+0.51} \pm 0.34 $ &   \\

$B^{0} \rightarrow K^{+} \pi^{0}$ & $32.3_{-8.4}^{+9.4}$ 
& $1.88_{-0.49}^{+0.55} \pm 0.23 $ &   \\

$B^{0} \rightarrow K^{0} \pi^{+}$ & $5.7_{-2.7}^{+3.4}$ 
& $1.66_{-0.78 - 0.24}^{+0.98 + 0.22}$  & $<$3.4   \\

$B^{0} \rightarrow K^{0} \pi^{0}$ & $10.8_{-4.0}^{+4.8}$ 
& $2.10_{-0.78 - 0.23}^{+0.93 + 0.25}$  &     \\ \hline

$B^{0} \rightarrow K^{+} K^{-}$ & $0.8_{-0.8}^{+3.1}$ 
&  & $<0.6$   \\

$B^{0} \rightarrow K^{+} K^{0}$ & $0.0_{-0.0}^{+0.5}$  
&  & $<0.51$   \\ \hline

$B^{0} \rightarrow \pi^{+} \pi^{-}$ & $9.3_{-5.1}^{+5.3}$ 
& $0.63_{-0.35}^{+0.39} \pm 0.16 $ & $<1.65$   \\

$B^{0} \rightarrow \pi^{+} \pi^{0}$ & $5.4_{-4.4}^{+5.7}$ 
& $0.33_{-0.27}^{+0.35} \pm 0.07 $ & $<1.01$  \\ \hline

\end{tabular}
\caption{Belle preliminary results for charmless  $B \rightarrow PP$ decays.
The first error is statistical, the second error is systematic. Upper limits
are given at the 90 \% C.L.}
\label{tab:B2PP}
\end{center}
\end{table}

\subsection{Radiative $B$ Meson Decays} 
Flavor-changing neutral decays involving $b \rightarrow s $ or
$b \rightarrow d$ transition have received much attention 
in recent years. The inclusive decay $B \rightarrow X_s \gamma$
where $X_s$ is a strange hadronic state, is of particular
interest to the experimentalist since the theoretical
description of the decay mode is rather clean and can be
related to the partonic weak decay $b \rightarrow s \gamma$.
A short term motivation in this direction was to measure
the branching fraction with a better particle identification
device and a high resolution electro-magnetic calorimeter.
Table 2 summarizes the signal yield and
corresponding branching fraction measurement 
of radiative $B$ meson decays at Belle. The results are comparable
to the recent CLEO results~\cite{cleo_radiative}. Our current
90 \% C.L. upper limit on the ratio is ${\cal B}(B \rightarrow \rho \gamma)/
{\cal B}(B \rightarrow K^{*} \gamma) <$0.28. This is an important result
because it constrains $|V_{td}/V_{ts}|$ within the Standard Model. 
A detailed description of the analysis method can be found 
in reference ~\cite{b2sgamma}.

\begin{table}[t]
\begin{center}
\begin{tabular}{llll}  
Decay Modes & Signal Yield  & ${\cal B}(\times 10^{-5})$ 
& U.L($\times 10^{-5}$) \\ \hline

$b \rightarrow s \gamma $ & $92 \pm 14$
& $33.4 \pm 5.0_{-0.37 - 2.8}^{+0.34 + 2.6} $ &   \\

$B^{0} \rightarrow K^{*0} \gamma $ & $33.7 \pm 6.9$ 
& $ 4.94 \pm 0.93_{-0.52}^{+0.55} $ & \\ 

$B^{+} \rightarrow K^{*+} \gamma $ & $8.7 \pm 4.2$ 
& $ 2.87 \pm 1.20_{-0.40}^{+0.55} $ & \\ 

$B^{0} \rightarrow \rho^{0} \gamma $ &  
&  & $<0.56$  \\

$B^{+} \rightarrow \rho^{+} \gamma $ &  
&  & $<2.27$  \\ \hline

\end{tabular}
\caption{Belle preliminary results for radiative $B$ meson decays.
The first error is statistical, the second error is systematic. Upper limits
are given at the 90 \% C.L.}
\label{tab:radiat}
\end{center}
\end{table}

\subsection{$B \rightarrow D^{(*)}K$}

We report the observation of the Cabibbo-suppressed decay modes 
$\bar{B}^{0} \rightarrow D^{*+} K^{-} $ and ${B}^{-} \rightarrow D^{*0} K^{-} $
(Fig. 2). In addition, we also report a new measurement of 
${\cal B}(\bar{B}^{+} \rightarrow \bar{D}^{0} K^{+})$. 
Thanks to the excellent particle identification at Belle,
one can clearly separate the signal from the background originating
from the Cabibbo-favored decay modes
with more than 3$\sigma$ significance.
We measured the ratio $R$ of the branching
fraction for the Cabibbo suppressed decay $B \rightarrow D^{(*)} K^{-}$
normalized relative to the Cabibbo allowed decay $B \rightarrow D^{(*)} \pi^{-}$.
The observed ratios are summarized
in Table~\ref{tab:BDK}. The detailed description of the analysis
method can be found in the reference ~\cite{b2dk}
As statistics improve, the 
analysis will shift towards the extraction of the CKM angle $\phi_3$.

\begin{table}[t]
\begin{center}
\begin{tabular}{ll}  
Decay Modes & Ratio \\ \hline

${\cal B}(B^{-} \rightarrow D^{0} K^{-}) / {\cal B}(B^{-} \rightarrow D^{0} \pi^{-})$
& $0.081 \pm 0.014 \pm 0.011$ \\

${\cal B}(B^{-} \rightarrow D^{*0} K^{-}) / {\cal B}(B^{-} \rightarrow D^{*0} \pi^{-})$
& $0.134_{-0.038}^{+0.045} \pm 0.015$ \\

${\cal B}(B^{-} \rightarrow D^{*+} K^{-}) / {\cal B}(B^{-} \rightarrow D^{*+} \pi^{-})$
& $0.062_{-0.024}^{+0.030} \pm 0.013$ \\ \hline

\end{tabular}
\caption{Belle preliminary results for Cabibbo-suppressed $B$ meson decays.
The first error is statistical, the second error is systematic.}
\label{tab:BDK}
\end{center}
\end{table}

\begin{figure}[htb]
\begin{center}
\epsfig{file=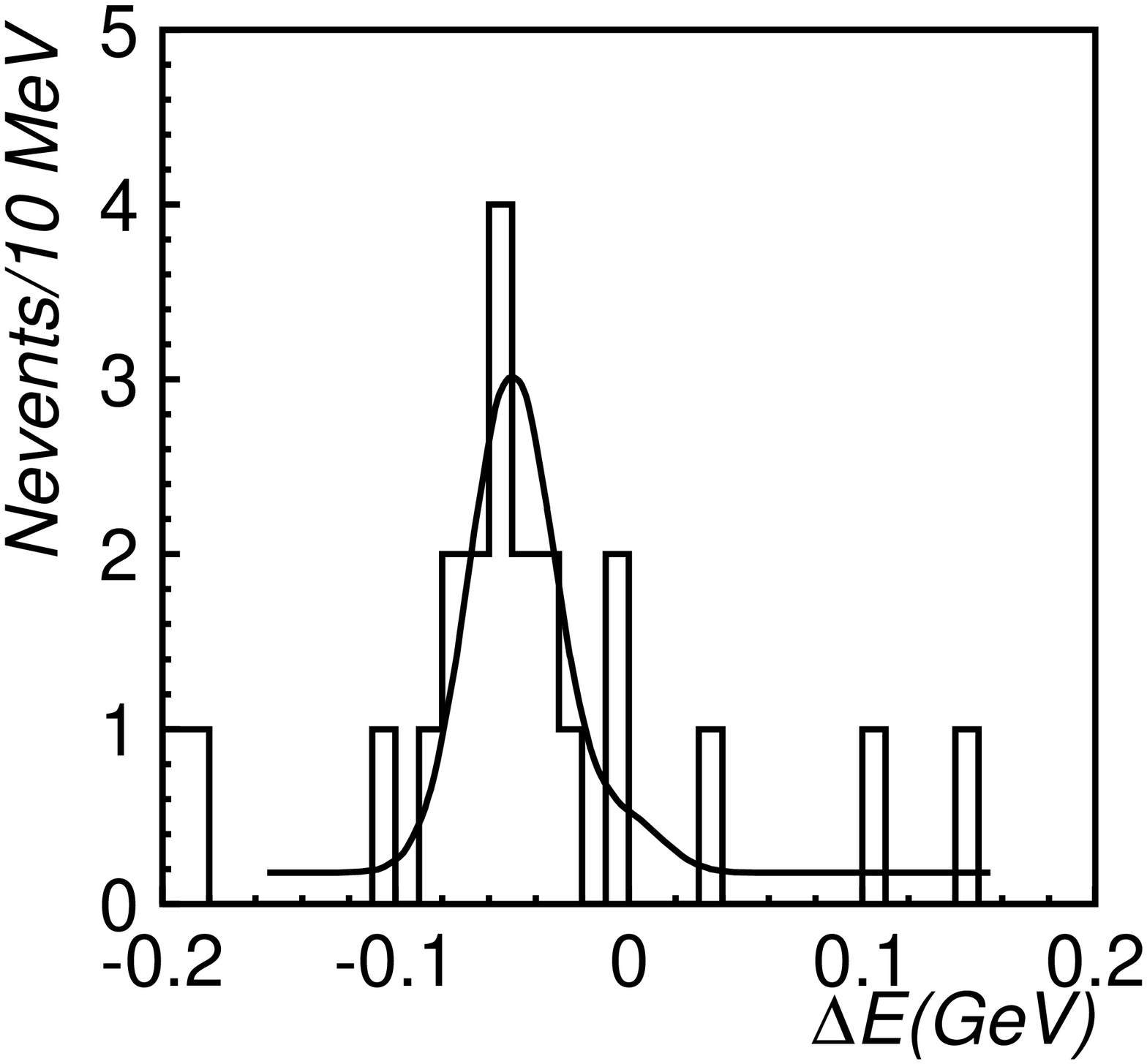,height=2.2in, width=2.2in}
\epsfig{file=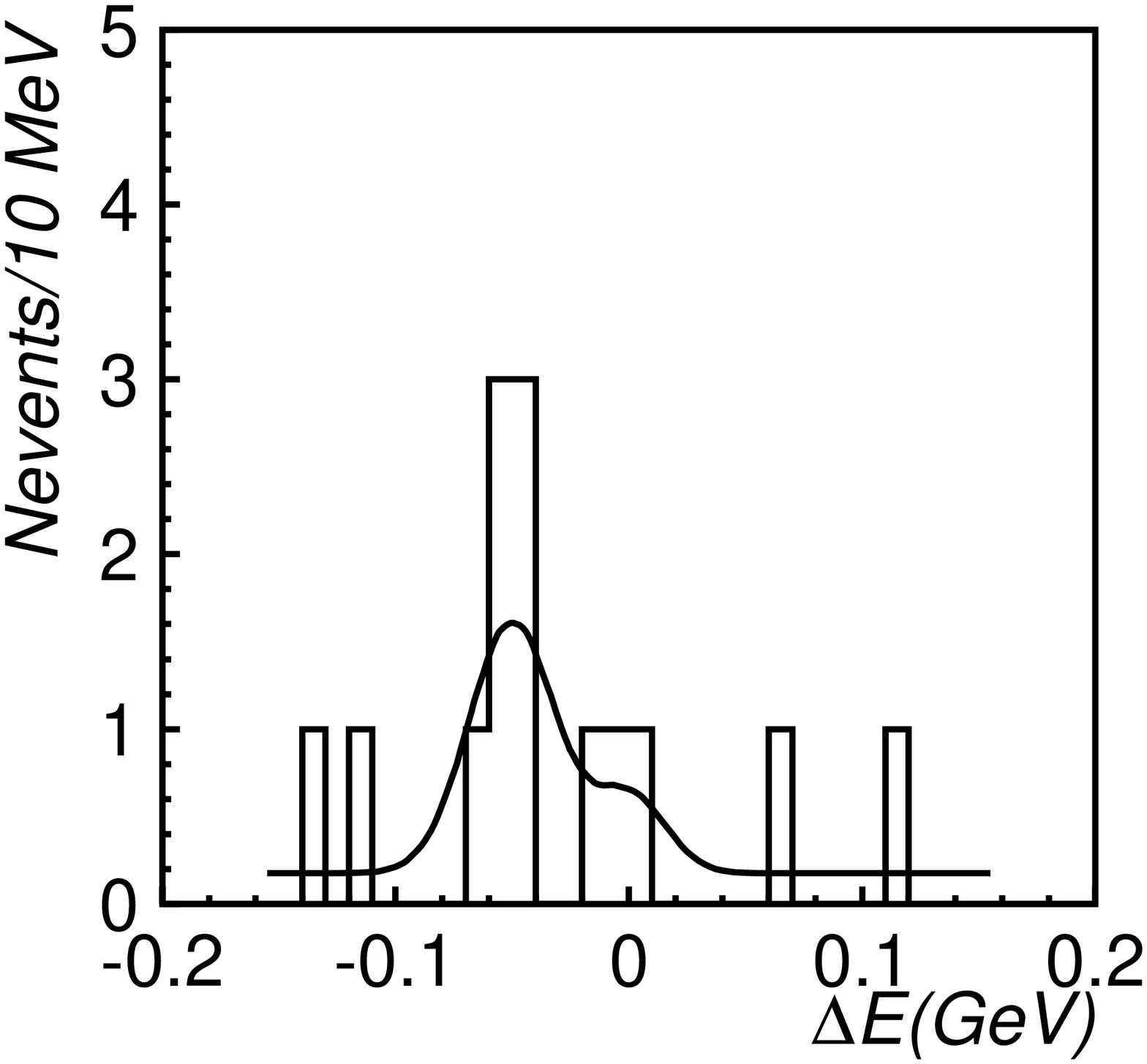,height=2.2in, width=2.2in}
\caption{$\Delta E $ distribution of 
(left) $B^{0} \rightarrow D^{*-} K^{+} $ 
(right) ${B}^{-} \rightarrow D^{*0} K^{-}$. The signal yield is 
obtained from the fit to the distribution with
a double Gaussian signal function and a MC determined
background shape.}
\label{fig:dsk}
\end{center}
\end{figure}

\subsection{$B \rightarrow J/\psi K_{1}$}

Inclusive $B \rightarrow J/\psi X$ decays are not saturated 
by the sum of observed 
exclusive modes. This motivates the search for 
new exclusive modes that we are reporting here.
$K_1$ candidates were reconstructed from 
$K^{+} \pi^{+} \pi^{-}, K^{+} \pi^{-} \pi^{0}$
and $K^{0} \pi^{+} \pi^{-}$. We have verified the
signal is due to $B \rightarrow J/\psi K_{1}(1270) $ (Fig. 3) 
and determine the branching fractions as summarized in Table~\ref{tab:BJK1}.

\begin{table}[t]
\begin{center}
\begin{tabular}{ll}  
Decay Modes & Branching Ratio ($ \times 10^3$) \\ \hline
${\cal B}(B^{0} \rightarrow J/\psi K_{1}^{0}(1270))$ 
& $1.5_{-0.4}^{+0.5} \pm 0.4 $ \\
${\cal B}(B^{+} \rightarrow J/\psi K_{1}^{+}(1270))$ 
& $1.7_{-0.4}^{+0.5} \pm 0.4$ \\ \hline
\end{tabular}
\caption {Belle preliminary results for $B \rightarrow J/\psi K_{1}(1270)$.
The first error is statistical, the second error is systematic.}
\label{tab:BJK1}
\end{center}
\end{table}

\begin{figure}[htb]
\begin{center}
\epsfig{file=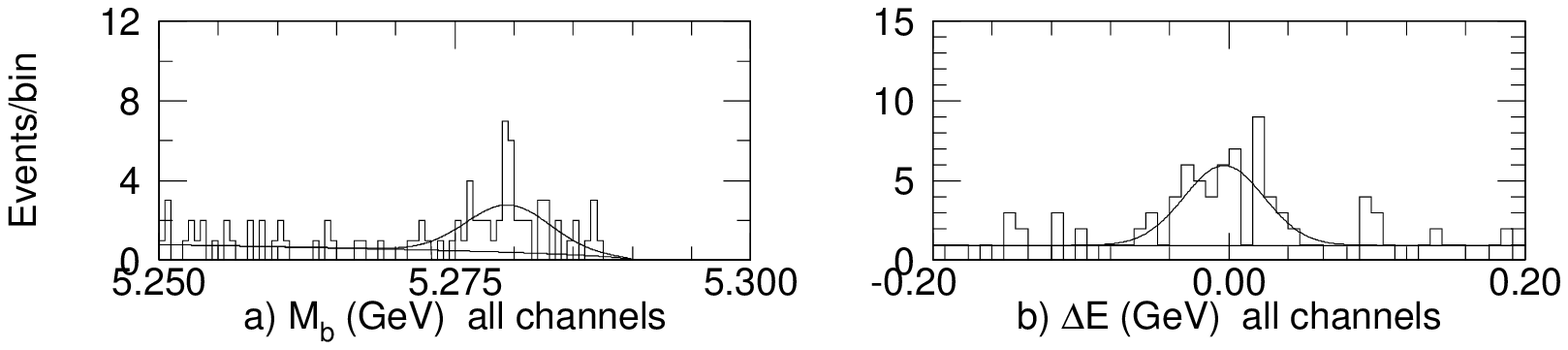, height=2.2in, width=4.9in}
\caption{Signal yield for the decay $B \rightarrow J/\psi K_{1}(1270)$}
(left) $M_{b}$ distribution 
(right) $\Delta E$ distribution.
\label{fig:jpsik1}
\end{center}
\end{figure}

\section{Measurement of $\sin(2\phi_{1})$ }

Experimentally, $CP$ asymmetry is observed in the distribution
of the proper time difference of two $B$ decays produced in pairs
in the decays of the $\Upsilon(4S)$, one to $CP$ eigenstate and 
another to any final state where the flavor is identified.
For a $B$ decaying to a $CP$ eigenstate, the time dependent asymmetry 
a(t) can be non-zero, indicating $CP$ violation :
\begin{equation}
a(t) = \frac{N(B^{0}(t) ~\rightarrow f) - N(\bar{B^{0}}(t) ~\rightarrow f)}
{N(B^{0}(t) ~\rightarrow f) + N(\bar{B^{0}}(t) ~\rightarrow f)} =
\frac{(1-|\lambda_f|^{2}) \cos(\Delta m_d t) - 2 Im \lambda_f \sin(\Delta m_d t)}
{(1+|\lambda_f|^2)} 
\end{equation}
where $\lambda_f = \frac{q}{p} \frac{A(\bar{B^{0}} \rightarrow f)}{A(B^{0} 
\rightarrow f)}$, $\Delta m_d$ is the $B_d$ mixing frequency, $\Gamma$ its width and
$Im \lambda_f = \sin(2\phi_1)$ that arises from the interference 
between the decays with and without mixing.

When $B^{0} \rightarrow f$ = $\bar{B^{0}} \rightarrow f$ and assuming only one 
diagram dominates the decay process, $|\lambda_f|^{2}$=1. Then the 
time dependent asymmetry would be
\begin{equation}
a(t) \sim \eta_{CP} \sin(2\phi_1) ~\sin(\Delta m_d t)
\end{equation}
where $\eta_{CP}$ = -1 for $\psi K_s$ type modes and $\eta_{CP}$ = 1 for $J/\psi K_L$
and $\psi \pi^{0}$ modes.
One of the important and immediate goal for the Belle experiment is to 
measure $\sin(2\phi_1)$ to see if the CKM model is correct.

\subsection{Event Reconstruction : $B$ Decaying to $CP$ Eigenstate}
\begin{figure}[htb]
\begin{center}
\epsfig{file=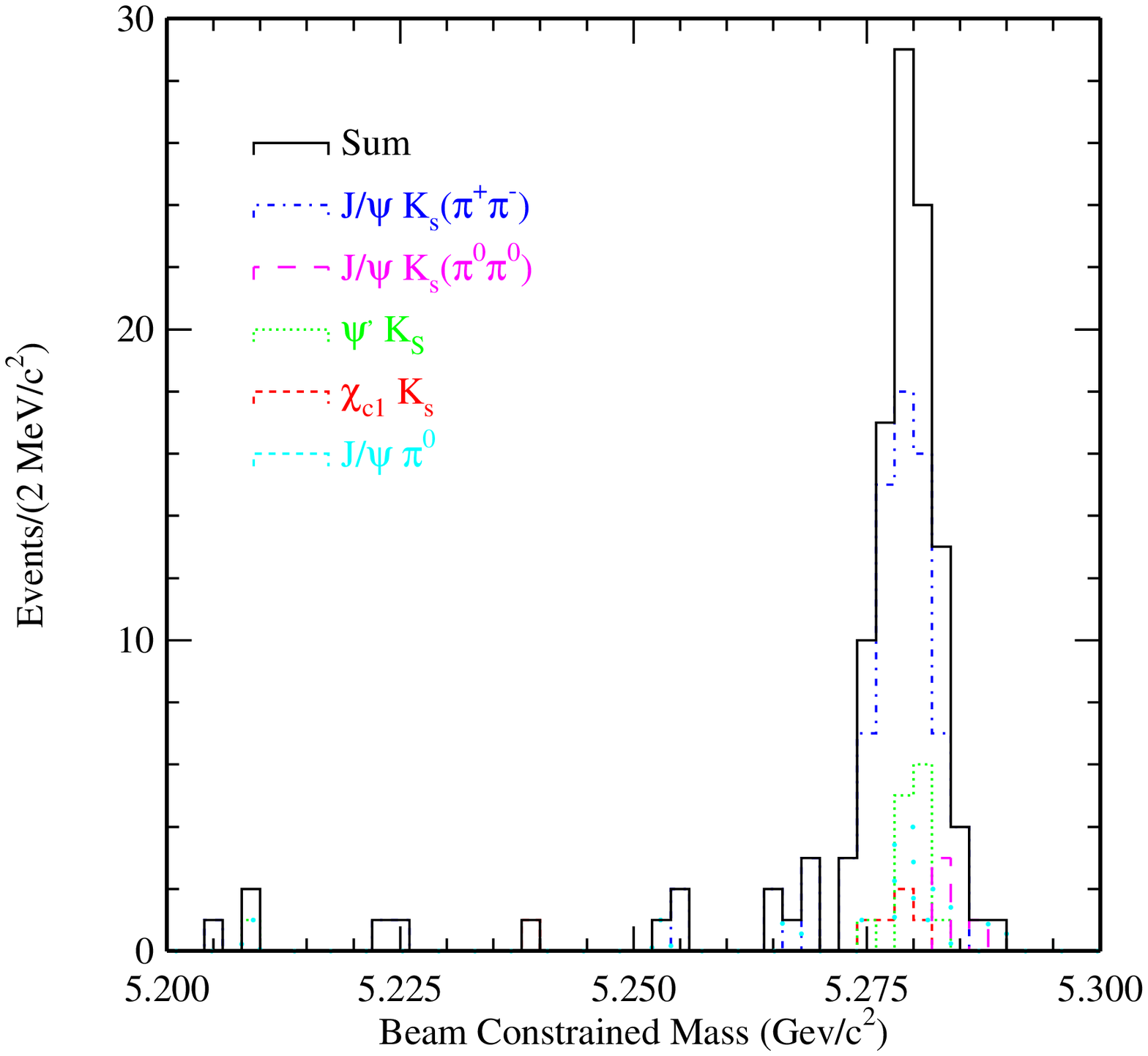,height=2.2in, width=2.2in}
\epsfig{file=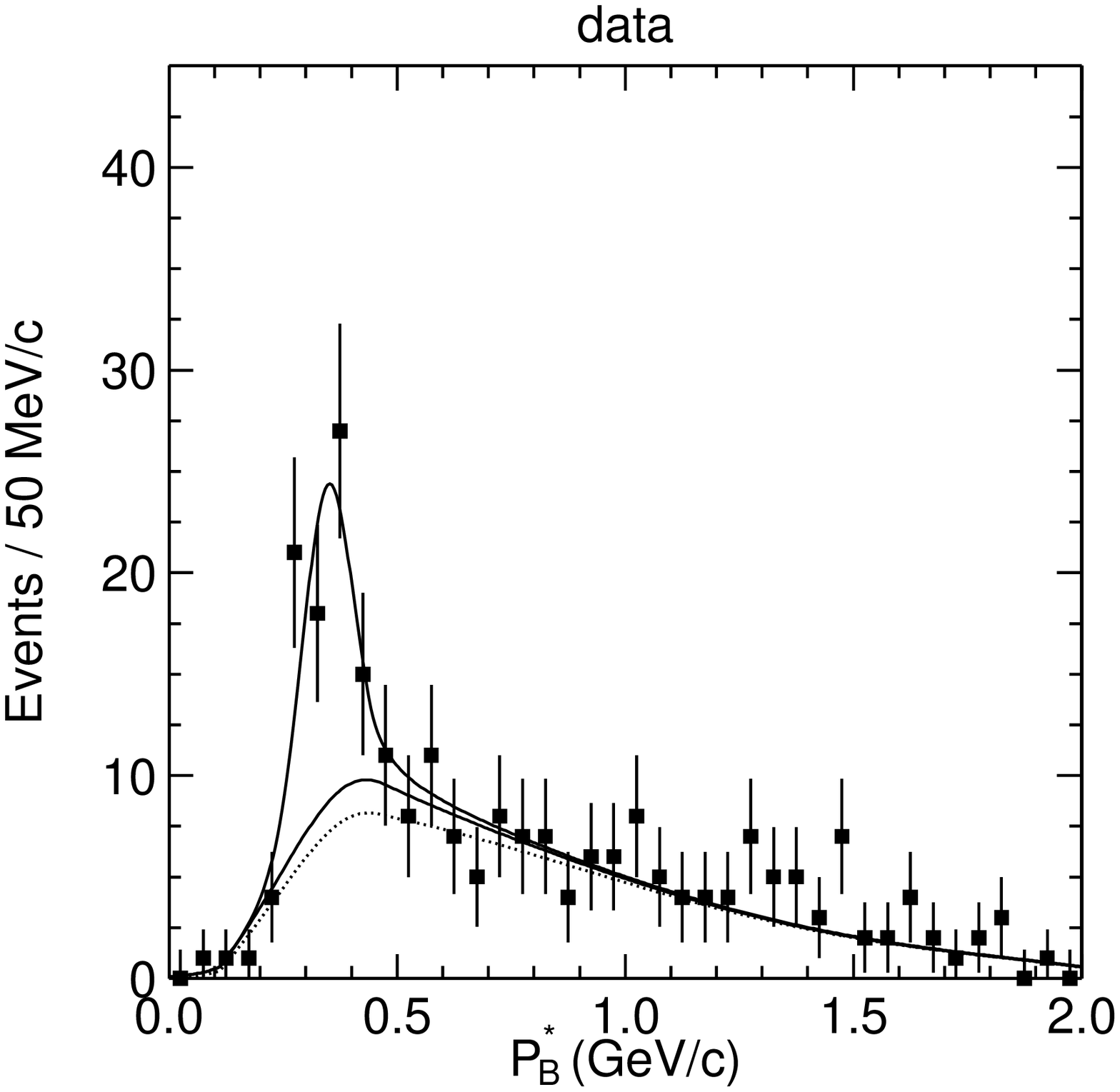,height=2.3in, width=2.2in}
\caption{$B_{CP}$ signal yield. [left:] Beam constrained
mass distribution of all the decays modes except the $K_{L}$ channel.
[right:] Momentum distribution of $B$ candidate events in CMS
where $B \rightarrow J/\psi K_{L}$ }
\label{fig:Brec}
\end{center}
\end{figure}

Table~\ref{tab:Bcp} summarizes the decay modes that are reconstructed for 
the CP analysis. $J/\psi$ and $\psi(2S)$ candidate events were
reconstructed from dileptons ($\mu^{+} \mu^{-}, e^{+} e^{-}$),
correcting for the final state radiation in the electron channel.
For $\psi(2S)$ candidates we also used the $J/\psi \pi^{+} \pi^{-}$ mode.
$\chi_{c1}$ candidates were reconstructed using only the $J/\psi \gamma $ 
decay mode. $K_s$ candidates were reconstructed in  
the $\pi^{+} \pi^{-}$ and $\pi^{0} \pi^{0}$ modes.
We reconstructed 102 $CP$ eigenstate candidate with 6 
estimated background (Fig.~\ref{fig:Brec}(left)). 

We also reconstructed 102 $CP$ even 
$B \rightarrow J/\psi K_{L}$ candidate events 
(Fig.~\ref{fig:Brec}(right)) with
48 estimated background. Among the backgrounds which contain
$CP$ asymmetry, major contributions come from  physics
events such as $B$ decays to $\chi_{c1}K_L$, $J/\psi K_s$,
$J/\psi K^{*}$ and  $J/\psi$ non-resonant $K_L \pi^0$.

\begin{table}[t]
\begin{center}
\begin{tabular}{llll}  
 Modes   & CP   & S / N  & Tagged\\ \hline
 $J/\psi(l^+ l^-) K_s(\pi^+ \pi^-)$   & -1    & 70 / 3.4 & 40 \\ 
 $J/\psi(l^+ l^-) K_s(\pi^0 \pi^0)$   & -1    & 4 / 0.3 & 4 \\ 
 $\psi^{\prime}(l^+ l^-) K_s(\pi^+ \pi^-)$ & -1   & 5 / 0.2 & 2  \\ 
 $\psi^{\prime}(J/\psi \pi^+ \pi^-) K_s(\pi^+ \pi^-)$  & -1 & 8 / 0.6 & 3  \\ 
 $\chi_{c1}(J/\psi \gamma) K_s(\pi^+ \pi^-)$  & -1 & 5 / 0.75 & 3 \\ 
 $J/\psi(l^+ l^-) \pi^{0}$  & +1 & 10 / 1  & 4 \\ \hline 
 Total 		&    & 102 / 6.25  & 56 \\ \hline 
 $J/\psi(l^+ l^-) K_L$ & +1  & 102 / 48  & 42 \\ \hline

\end{tabular}
\caption{Summary of Signal Yield of CP eigenstate $B$ meson decays.}
\label{tab:Bcp}
\end{center}
\end{table}

\subsection{Measurement of $B$ Life Time : A Benchmark Test}

Extraction of $CP$ asymmetry requires the knowledge of
proper time distribution of tagged and fully reconstructed $B_{CP}$
event. For $CP$ eigenstate modes,
the proper time, $\Delta t = 
\Delta z /c \beta \gamma$ at $\beta \gamma = 0.425$,
was calculated by measuring the difference between the decay vertices of
$B_{CP}$ decay vertex and tagging side $B_{tag}$ vertex. The vertex point
of $B_{CP}$ was established by the two tracks associated with the $J/\psi$ 
decay. The vertex position in the tagging 
side was determined from the tracks not assigned to $B_{CP}$ by an
algorithm that removes tracks from the secondary vertices or 
tracks which makes a large increase in the $\chi^{2}$ of the vertex fit.

\begin{figure}[htb]
\begin{center}
\epsfig{file=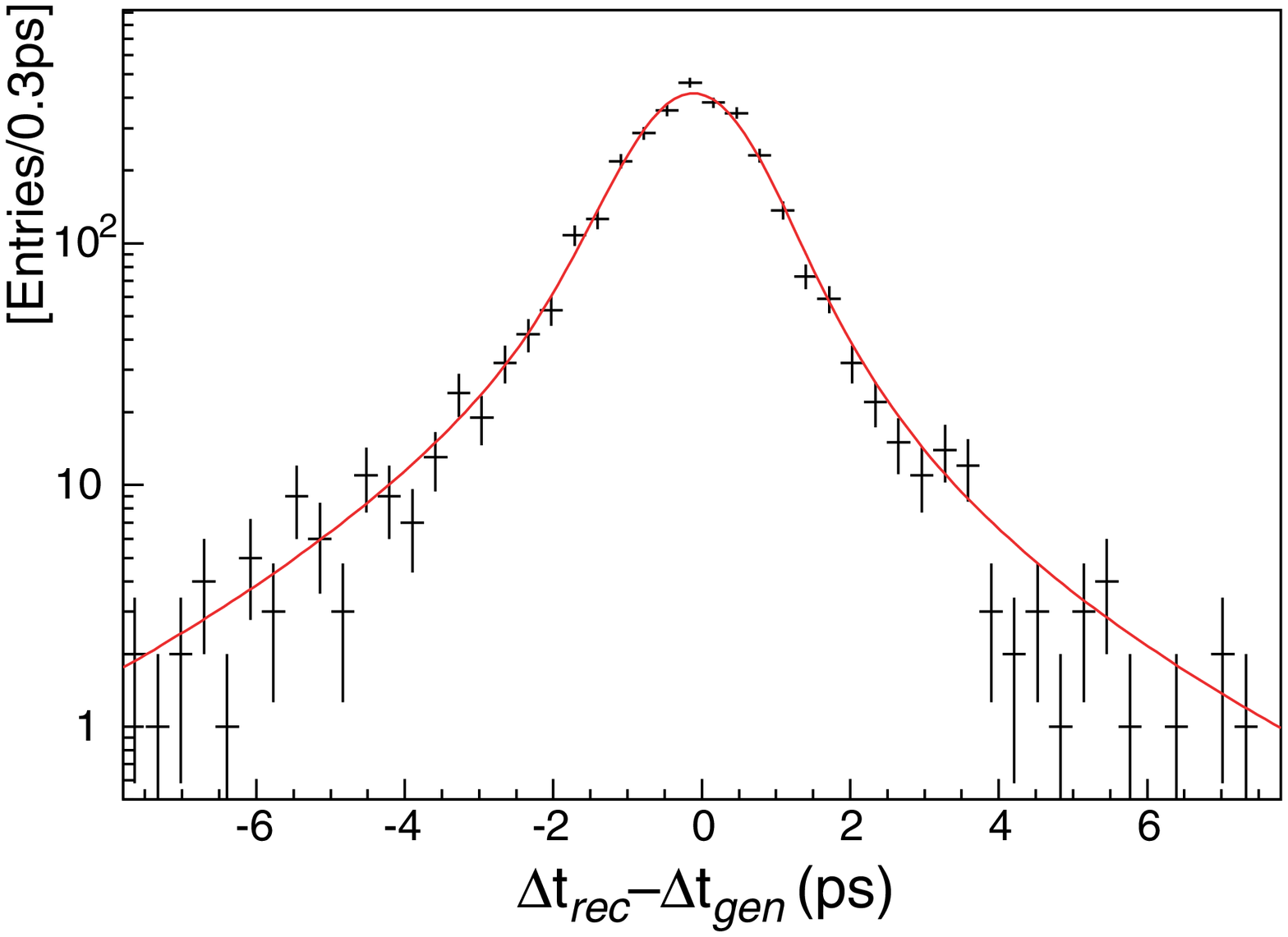,height=2.5in, width=2.2in}
\epsfig{file=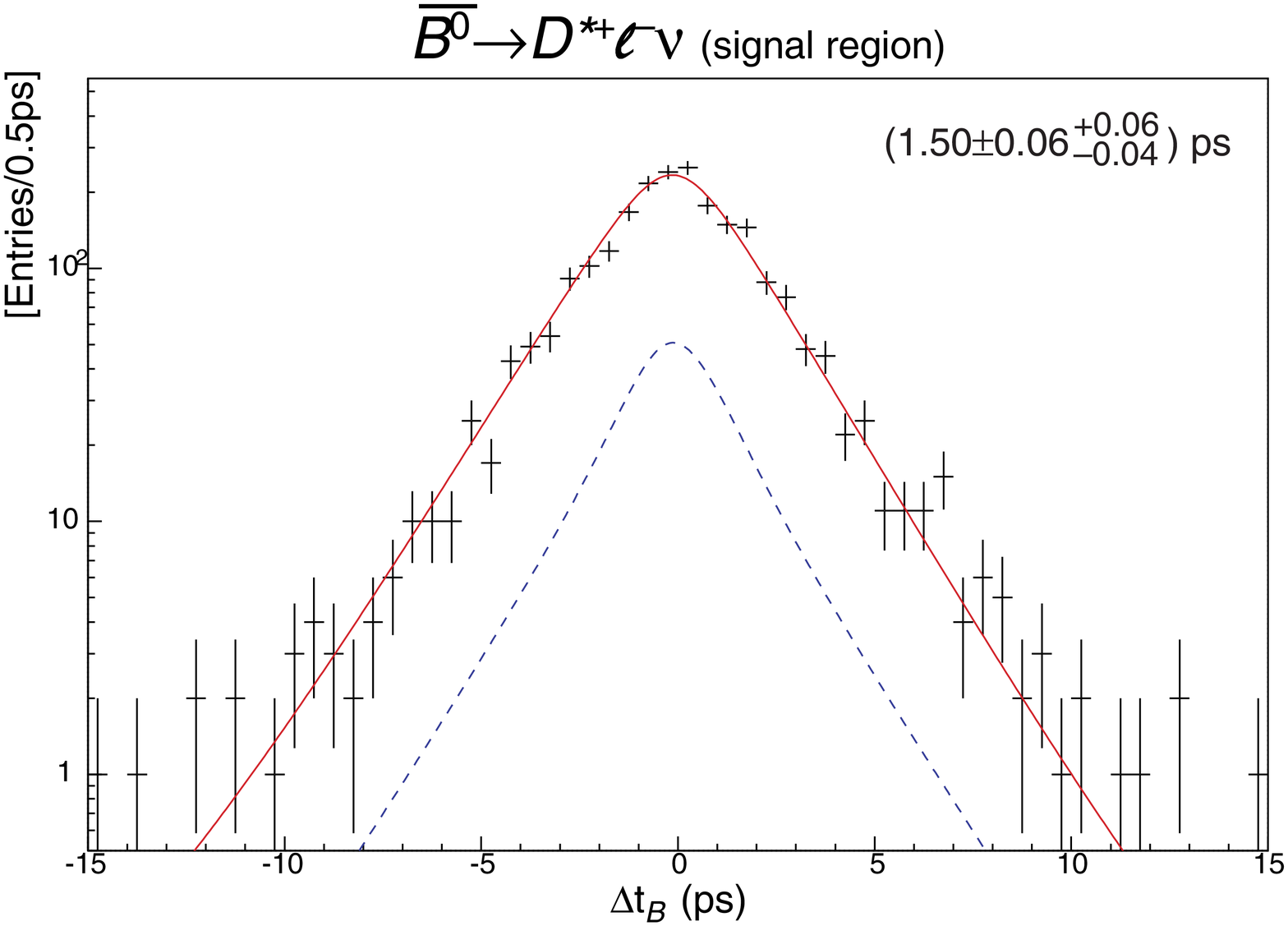,height=2.8in, width=2.4in}
\caption{(left):Average shape of the event by event resolution function.
(right): Lifetime fit results for $\bar{B^{0}} \rightarrow D^{*+}l^{-}\nu$.}
\label{fig:propertime}
\end{center}
\end{figure}

The proper time resolution function $R_{sig}(\Delta t)$ was parameterized
from MC simulation studies and
a multi-parameter fit to $B \rightarrow D^{*} l \nu$ data 
(Fig.~\ref{fig:propertime}).  
A double Gaussian parameterization 
results from various detector characteristics, error
in the event by event vertex fit, error in the $\Delta t$ approximation,
the scale factor and charm lifetime. 

We measured the $B$ life time in various decay modes to 
test the proper time resolution function that we derived from
MC studies.
The $B$ lifetime was extracted from an event by 
event likelihood fit with a P.D.F given by	
\begin{eqnarray}
P(\Delta t) &=& f_{sig} \int_{-\infty}^{\infty} d(\Delta t^{\prime})
\frac{e^{-|\Delta t^{\prime}|/ \tau_{sig}}}{2 \tau_{sig}}	
R_{sig}(\Delta t - \Delta t^{\prime}) \nonumber \\
& & +(1-f_{sig}) \int_{-\infty}^{\infty}d(\Delta t^{\prime}) 
[f_{\lambda_{bg}} 
\frac{\lambda_{bg}}{2} e^{-|\Delta t^{\prime}|\lambda_{bg}} + 
(1-f_{\lambda_{bg}}) \delta(\Delta t^{\prime}) ] 
R_{bg} (\Delta t - \Delta t^{\prime})
\end{eqnarray}
Minimization of the likelihood gives $\tau_{B}$ which is one of the  
free parameters determined from the fit. It should be noted that
$R_{bg}(\Delta t)$ was parameterized in the same way as of $R_{sig}(\Delta t)$
using background events from the sideband region of the $\Delta E$ and $M_{b}$ 
scatter plot. 
 
The superimposed solid line on the data
points in the right plot of
Fig.~\ref{fig:propertime} is the result of
the refit for $\bar{B^{0}} \rightarrow D^{*+} l^{-} \nu$.
The measured $\tau_{B}$ agrees with the world average value
proving that the parameterization of the resolution function is correct.
Table~\ref{tab:Blife} summarizes the measured combined results for the
$B$ lifetime from various charged and neutral $B$ decays.

\begin{table}[t]
\begin{center}
\begin{tabular}{lll}  
    & Belle   &Particle Data group  \\ \hline
 $\tau_{\bar{B^{0}}}$   & $1.50 \pm 0.05\pm 0.07$ ps & $1.548 \pm0.032$ ps \\ 
 $\tau_{B^{-}}$         & $1.70 \pm 0.06_{-0.10}^{+0.11}$ ps & $1.653 \pm 0.028$ ps \\ 
 $\tau_{B^{-}}/\tau_{\bar{B^{0}}}$ 
	                & $1.14 \pm 0.06_{-0.05}^{+0.06}$    & $1.062 \pm 0.029$ \\ \hline
\end{tabular}
\caption{Summary of the measured $B$ lifetime at Belle.}
\label{tab:Blife}
\end{center}
\end{table}

\subsection{Flavor Tagging and Wrong Tag Fraction}
To measure the $CP$ asymmetry, we need to
determine the flavor of the $B_{CP}$ candidates from the 
remaining tracks of the event.
We use the following algorithm in sequence to tag a certain $B$ in an event.
(0) Require tight PID probability for lepton and kaon selection,
(1) Obtain the sign of the high momentum lepton ($p^{*}_{l} > 1.1$ GeV) 
$\rightarrow$ positive lepton tags $B^{0}$,
(2) Obtain sum of $K$ charges $\rightarrow$ positive sum tags $B^{0}$,
(3) Look for medium momentum leptons ($0.6 <p^{*}_{l} <1.1$ GeV) and large missing
momenta $\rightarrow$ positive charged lepton tags $B^{0}$,
(4) Sign of slow pions from $D^{*+}$ decays $\rightarrow$
charge of slow pion and hence flavor of $D^{*}$ tags the flavor of $B$.

The algorithm was tested on a sample of self tagging
exclusively reconstructed $B \rightarrow D^{(*)} l \nu$ decays.
We extract the wrong tag fraction ($w$) and $\Delta m_d$
from a maximum likelihood fit to the $\Delta t$ distribution (eqn. 4)
of $OF$ (opposite flavor) and $SF$ (same flavor)
events with a function that includes
the effect of $\Delta t$ resolution and background.

\begin{equation}
A_{mix}(\Delta t) 
= \frac{N(\Delta t)^{OF}-{N(\Delta t)^{SF}}}{N(\Delta t)^{OF}+
{N(\Delta t)^{SF}}} = (1-2w)\cos(\Delta m_d t)
\end{equation} 	

The measured value $\Delta m_{d} = 0.488 \pm 0.026$ (stat.) ps$^{-1}$ 
verifies the consistency of the tagging algorithm. In an independent
approach, $\Delta m_{d}$ was determined from the time evolution of
dilepton yields in $\Upsilon(4S)$ decays. The proper-time 
difference distribution for same-sign and opposite-sign dilepton
events were simultaneously 
fitted to an expression containing  $\Delta m_{d}$ as a free parameter.
Using both electrons and muons, we obtain 
$\Delta m_{d} = 0.463 \pm 0.008$ (stat.)$\pm 0.016$ (sys) ps$^{-1}$
~\cite{mixing_dilepton}.
This is the first determination of $\Delta m_d$ from time
evolution measurements at the $\Upsilon(4S)$. Previous measurements
at the $\Upsilon(4S)$ only used time integrated distributions.

Table~\ref{tab:tageff} summarizes the estimated
tagging efficiency in each of the above steps. The effective 
tagging efficiency $\epsilon_{eff} = \epsilon_{tag} (1-2w)^2$ is 
found to be 21.2 \%.
We also determined the tagging efficiency from the same test sample.
Depending on the tag type, we used the numbers in the table for the fit to
extract $\sin(2\phi_{1})$. 
\begin{table}[t]
\begin{center}
\begin{tabular}{llll}  
Method             &$\epsilon_{tag}$    & $w$ (\%)     & $\epsilon_{eff}$ \\ \hline
High $p^*$ Lepton  & 14.2$\pm$2.1       &7.1$\pm$4.5   & 10.5$\pm$2.7     \\ 
Kaons              & 27.9$\pm$4.2       &19.9$\pm$7.0  & 10.1$\pm$4.9     \\ 
Mid $p^*$ Lepton   & 2.9$\pm$           &29.2$\pm$15.0 & 0.5		  \\ 
Soft pion          & 7.0$\pm$3.5        &34.1$\pm$15.0 & 0.7		  \\ \hline 
Total              & 52.0               &              & 21.2 	          \\ \hline
\end{tabular}
\caption{Tagging efficiency and wrong tagging fraction at Belle}
\label{tab:tageff}
\end{center}
\end{table}

\subsection{CP Fit :  Extraction of $\sin(2\phi_{1})$ }
Using the tagging method described above, from a sample of 102 $CP$ odd and $J/\psi
\pi^{0}$ events and 102 $J/\psi K_{L}$ events, a total of 98 events
were tagged. The likelihood for each tagged event is calculated as
\begin{eqnarray}
P(\Delta t) & = & 
f_{sig}\int_{-\infty}^{\infty} Sig(\Delta t^{\prime}, \eta_{CP})
R_{sig}(\Delta t - \Delta t^{\prime}) d(\Delta t^{\prime}) \nonumber \\
& & + (1-f_{sig})\int_{-\infty}^{\infty}Bkg(\Delta t^{\prime})
R_{sig}(\Delta t - \Delta t^{\prime}) d(\Delta t^{\prime})
\end{eqnarray}
where the P.D.F expected for the signal distribution with $CP$ eigenvalue
$\eta_{CP}$ is: \\
$Sig(\Delta t, \eta_{CP}) = 
\frac{1}{\tau_{B^{0}}} \exp(-|\Delta t|/\tau_{B^{0}})
 \{1\mp\eta_{CP}(1-2w)\sin(2\phi_1)\sin(\Delta m_d \Delta t)\}$ \\ and
that of background distribution is:
$Bkg(\Delta t^{\prime})= \frac{1}{2\tau_{bkg}} \exp(-|\Delta t/\tau_{bkg})$
and $w$ depends on the method of flavor tagging for each event.

\begin{figure}[htb]
\begin{center}
\epsfig{file=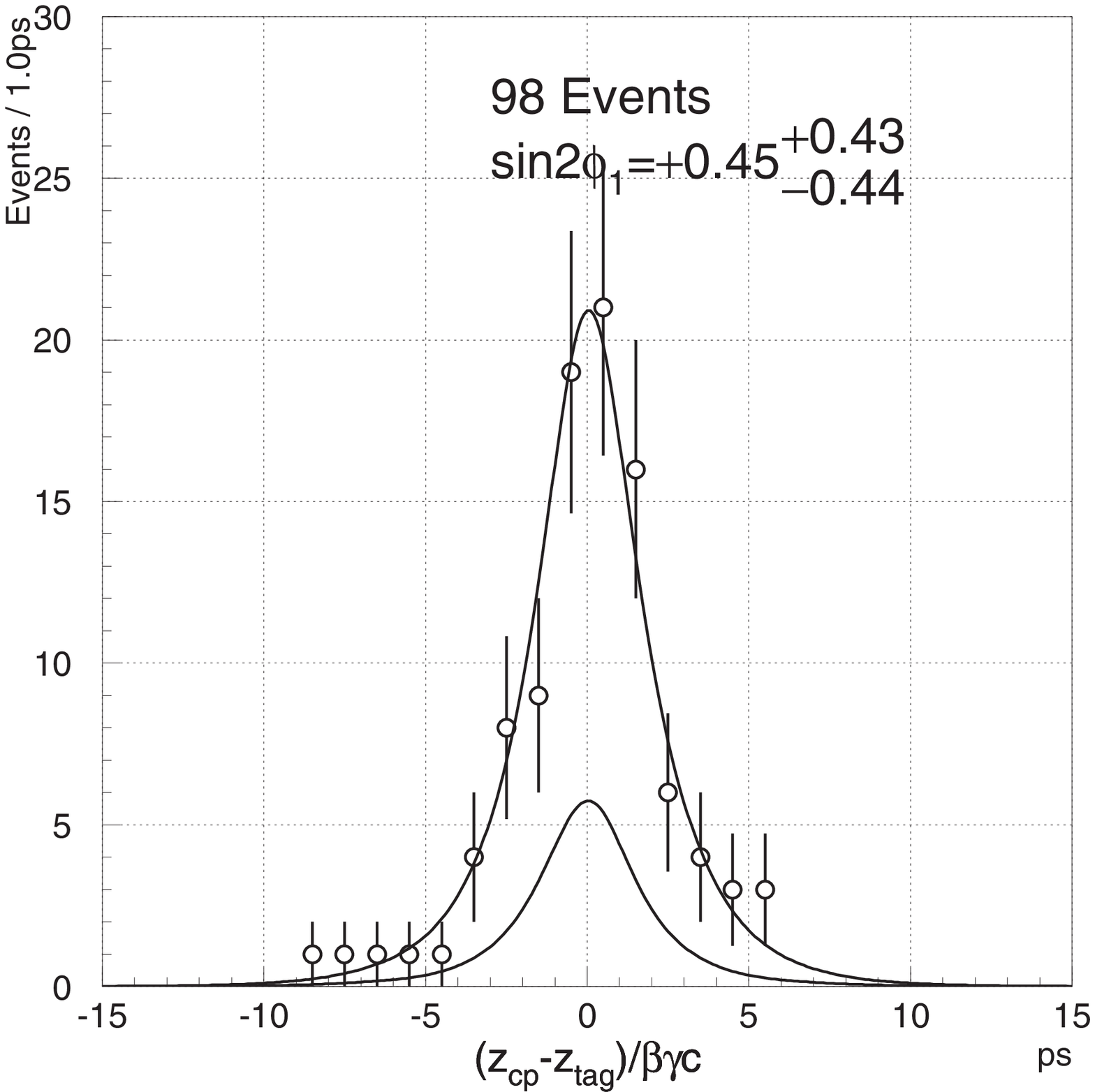,height=2.2in, width=2.2in}
\epsfig{file=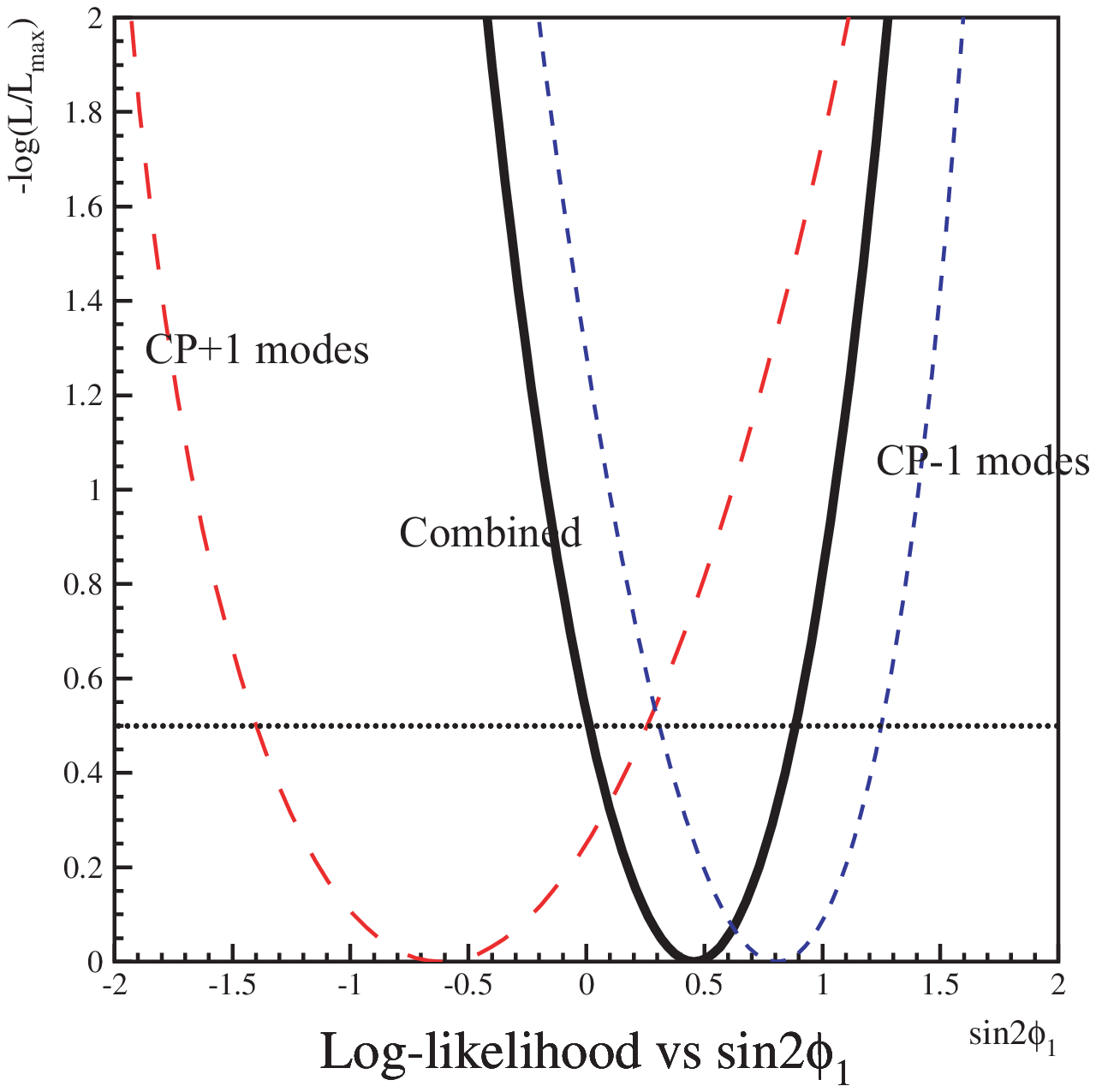,height=2.2in, width=2.2in}
\caption{CP fit results for a combined $CP$ even and $CP$ odd events.}
\label{fig:cpfit}
\end{center}
\end{figure}

The values of $\Delta m_d$ and $\tau_{B^{0}}$ are
fixed to the ones in the P.D.G.	
We use an unbinned maximum 
likelihood fit to extract the possible $CP$ asymmetry.
Before doing a $CP$ fit, we wanted to make sure that the whole fitting procedure
is bias free. We performed the same analysis procedure
including tagging to several non-$CP$ eigenstate decay modes,
such as 
$B^0 \rightarrow J/\psi K^{*0}$, $B^- \rightarrow J/\psi K^- , D^0 \pi^- $
and found $\sin(2\phi_{1})$ is
consistent with zero within fitting errors.
Our result from the 
likelihood fit to the fully tagged sample is :
$\sin(2\phi_1) = +0.45_{-0.44}^{+0.43}(stat.)_{-0.09}^{+0.07}(syst.)$
(Fig.~\ref{fig:cpfit}).Clearly the measurement is statistics limited. 
It should be noted that the uncertainty
in determining the wrong tag fraction is the largest contribution
in the systematics.

\section{Conclusion and Prospect}
Belle had a very successful and exciting first year run and is 
marching along a well defined road to measuring $\sin(2\phi_1)$ with
a very good precision. First preliminary results were reported.
We need a lot more data to constrain the CKM triangle with small errors. 
We observed the first evidence of 
Cabibbo suppressed $B \rightarrow D^{*} K^-$ process,
and made first measurements of 
${\cal B}(B \rightarrow J/\psi K_1(1270))$ and 
${\cal B}(B^{+} \rightarrow \phi K^{+}$).
New results on many rare decays which will be used to search for
direct $CP$ violation have also been reported.
The physics scope at Belle is not limited to $B$ physics only.
We have reported five different $\tau$ and two-photon physics related 
results at the ICHEP2000 conference. Please check out for more:
http://www.bsunsrv1.kek.jp/conferences/ichep2000.html.

\Acknowledgements
We gratefully acknowledge the effort of the KEK-B staff in providing 
us with excellent luminosity and running condition. We thank the rest
of the Belle Collaboration for their support and help. Special thanks
to the KEK computing research center for their help in many different
occasion.

\end{document}